\documentclass[twocolumn,showkeys,amsmath,amssymb]{revtex4-1}
\usepackage[T1]{fontenc}
\usepackage{color}
\usepackage[latin1]{inputenc}
\usepackage{graphicx}
\usepackage{epsfig}
\usepackage{rotating}
\usepackage{dcolumn}
\usepackage{bm}
\usepackage{mathtools}

\begin{document}

\title{Ordering of hard rectangles in strong confinement}

\author{P\'eter Gurin$^a$, Szabolcs Varga$^a$, Miguel Gonz\'alez-Pinto$^b$, 
Yuri Mart\'{\i}nez-Rat\'on$^c$ and Enrique Velasco$^b$}
\affiliation{$^a$
Institute of Physics and Mechatronics, University of Pannonia, PO Box 158, Veszpr\'em, H-
8201 Hungary\\
$^b$ Departamento de F\'{\i}sica Te\'orica de la Materia Condensada, Instituto de F\'{\i}sica de la
Materia Condensada (IFIMAC) and Instituto de Ciencia de Materiales Nicol\'as Cabrera,  Universidad Aut\'onoma de Madrid, E-28049 Madrid, Spain\\
$^c$ Grupo Interdisciplinar de Sistemas Complejos (GISC), Departamento de Matem\'aticas,
Escuela Polit\'ecnica Superior,  Carlos III de Madrid, Avenida de la Universidad
30, E-28911, Legan\'es, Madrid, Spain
}

\date{\today}

\begin{abstract}
  Using transfer operator and fundamental measure theories, we examine
  the structural and thermodynamic properties of hard rectangles
  confined between two parallel hard walls.  The side lengths of the
  rectangle ($L$ and $D$, $L>D$) and the pore width ($H$) are chosen
  such that maximum two layers are allowed to form in planar order
  ($L$ is parallel to the wall), while only one in homeotropic order
  ($D$ is parallel to the wall). We observe three different
  structures: (i) a low density fluid phase with parallel alignment to
  the wall, (ii) an intermediate and high density fluid phase with two
  layers and planar ordering and (iii) a dense single fluid layer with
  homeotropic ordering. The appearance of these phases and the change
  in the ordering direction with density is a consequence of the
  varying close packing structures with $L$ and $H$.  Interestingly,
  even three different structures can be observed with increasing
  density if $L$ is close to $H$.
\end{abstract}


\maketitle

     \section{Introduction}

     The properties of molecular and colloidal systems can be altered
     substantially in restricted geometries such as slit-like pores,
     cylindrical tubes and spherical cavities. Even the simple hard
     sphere system confined between two parallel hard walls exhibits
     rich phase behaviour with changing the wall separation \cite{1}.
     Due to the commensuration effect between the size of the sphere
     and the wall separation several intermediate (e.g. prism and
     rhombic phases) and crystalline structures with layers of square
     and triangle symmetries can be generated \cite{2}. However, the
     phase behaviour of non-spherical hard bodies is even richer both
     in bulk and confinement as several liquid crystalline structures
     (e.g. nematic and smectic A), director distortion and domain
     walls may emerge due to the orientation dependence of
     particle-particle and wall-particle interactions. The ordering of
     hard rods between two planar hard walls is versatile as planar
     ordering and surface induced biaxial order may emerge in the
     vicinity of walls. As hard walls promote the orientational
     ordering, the isotropic phase may be suppressed and the
     isotropic-nematic transition (capillary nematization) terminates
     at a critical point with decreasing wall separation
     \cite{3,4,5,6,7}. In addition to this the nematic-smectic A
     transition can be also suppressed and even layering transitions
     between periodic structures with $n$ and $n+1$ layers can be induced
     \cite{8,9}.

     The system of two-dimensional (2D) hard objects, which can be
     realized by strong confinement of three-dimensional colloidal
     particles into a plane, can also exhibit wealthy phase behaviour
     in bulk and confinement. Among these the monolayer of square and
     rectangle-shaped hard particles has been studied extensively with
     simulation \cite{10,11,12}, theory \cite{13,14,15,16,17,18,19}
     and experiment \cite{20,21,22,23,24,25}. Interestingly, a
     tetratic phase of four-fold symmetry can be observed both in hard
     square \cite{10} and weakly elongated hard rectangle systems
     \cite{11}, while a nematic phase of two-fold symmetry can be
     stabilized in the system of elongated hard rectangles \cite{13}.
     Several studies are devoted to the effect of confinement on the
     phase behaviour of hard squares and rectangles. In spherical and
     square cavities the nature of wall induced defective structures
     (e.g. topology of the defects, domain formation) are examined by
     density functional theories \cite{26,27,28,29} and Monte Carlo
     simulations \cite{29,30,31,32}.  In slit pores the capillary
     nematisation and layering transitions are observed for large
     shape anisotropies with walls favouring homeotropic anchoring
     \cite{32}. In the case of hard walls, planar ordering and
     increased nematic ordering can be seen with respect to isotropic
     and tetratic order \cite{33,34}.

     Here we study the system of hard rectangles in a very narrow
     slit-like pore and search for the possibility of surface
     ordering, the formation of different structures and structural
     transitions. With the help of the transfer operator method (TOM) we
     obtain exact results for the thermodynamic quantities (e.g.
     density, heat capacity, vertical or normal pressure), while the
     fundamental measure density functional theory (FMT) is applied to
     get further information about the structure of observed phases.
     As the TOM of classical fluids is originally devised for one
     dimensional fluids \cite{35,36,37}, we extend the method for
     confinements such that a maximum of two layers can form in planar
     ordering (particles' long axes are parallel to the wall) or only
     one layer can accommodate within the pore in homeotropic ordering
     (particles' long axes are perpendicular to the wall). We show
     that the structure of the fluid can be manipulated by the
     external longitudinal force, the particle's shape anisotropy and
     the width of the pore. The low density (small external force)
     structure is always dominated by planar ordering, while the high
     density (large external force) structures can exhibit both planar
     and homeotropic ordering with one or two fluid layers. It may
     happen that three different fluid structures emerge upon
     compression of the system at some particular shape anisotropy and
     pore-width.  However, our results clearly show that the changes
     in orientational and layering properties do not result in a true
     phase transition, but they correspond to structural transitions
     with marked peaks in heat capacity and compressibility.

     \section{Transfer operator method of confined hard rectangles}

     \begin{figure}
       \includegraphics[width=8cm]{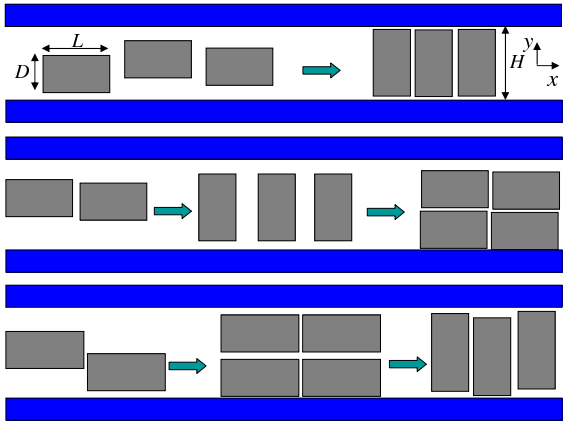}
       \caption{Possible phases and phase sequences of hard rectangles
         with increasing horizontal pressure ($P_x$) in a narrow hard
         channel, where $H$ is the width of the channel and $L$ ($D$)
         is the length (width) of the rectangles. At low pressures
         (densities) a fluid phase with one layer forms, where the
         particles align along the horizontal axis ($x$).  This phase
         is denoted as FH1.  At intermediate and high pressures (densities)
         either a single fluid layer with orientations along the
         vertical direction ($y$) or a fluid with two layers and
         horizontal order can be observed. These phases are referred
         to as FV and FH2, respectively. The upper panel shows a
         FH1-FV structural transition, while FH1-FV-FH2 and FH1-FH2-FV
         phase sequences can be seen in the middle and lower panels.}
       \label{fig1}
     \end{figure}

     Here we consider a system of $N$ hard rectangles of side lengths
     $L$ and $D$ in a narrow hard channel with pore width $H$ at a
     given external force acting along the channel.  The schematic of
     the channel and the possible structures of the rectangles are
     shown in Fig. \ref{fig1}.  The Cartesian coordinate system is
     chosen to be in the middle of the pore, where $x$ and $y$ axes
     are along the horizontal and vertical directions, respectively.
     The external force ($f_x$), which sets the density of the
     rectangles, is pointing along the $x$ axis. To get exact results
     for this system, we restrict our study to the following
     conditions for the pore width, molecular dimensions and
     orientations.  The allowed ranges of $L$ and $H$ are given by
     $D<L<3D$ and $L<H<\min\{L+D,3D\}$. Moreover, only horizontal and
     vertical orientations are allowed for the rectangles (restricted
     orientation model). In the horizontal state ($h$), the long axis
     of the particle is parallel to the confining walls, while the
     short axis of the particle is along the $x$ axis in the vertical
     state ($v$). These conditions guarantee that only particles being
     in $h$ states can pass each other, while there is no room for
     passage for $v$ state particles in the presence of $h$ or $v$
     state particle.  Therefore one or two layers can form in the
     horizontal state for $H>2D$, while only one in the vertical one
     (see Fig.~\ref{fig1}). We now extend the transfer operator
     method for rotating rectangles, which was developed for parallel
     hard squares \cite{38}. To do this, we start with the
     configurational part of the isobaric partition function of second
     neighbor interacting systems, which can be written as
\begin{align}
     Z
  =&\int_0^{\infty} dL_x e^{-\beta f_x L_x} \times
  \nonumber
 \\
   &\int \left( \prod_{i=1}^N dq_i \right) 
       e^{-\beta \sum_{j=1}^N [ u(q_j,q_{j+1}) + u(q_j,q_{j+2}) ] }
 ,
 \label{eq1}
\end{align}
     where $\beta=1/k_BT$ is the inverse temperature, $L_x$ is the
     length of the pore, $q_i=(x_i,y_i,\varphi_i)$ is a notation of
     the position and orientation of particle $i$ and $u(q_i,q_j)$ is
     the pair potential between particles $i$ and $j$.  Note that
     $1/N!$ is missing, because we tacitly assumed that
     $0<x_1<x_2<\cdots<x_N<L_x$. We also employ the periodic boundary
     condition, which means that $q_{N+1}=q_1$ and $q_{N+2}=q_{2}$. The
     ranges of the integrals for the $y$ positions in Eq.~(\ref{eq1})
     are $-(H-\sigma(\varphi_i))/2<y_i<(H-\sigma(\varphi_i))/2$.  In
     our two-state model the orientation can be either horizontal or
     vertical, i.e. $\varphi_i=\{h,v\}$, and the vertical length of
     the particle is given by $\sigma(h)=D$ and $\sigma(v)=L$.
     Furthermore, the integral $\int d\varphi$ which is included in
     the $\int dq$ notation, is understood as a sum,
     $\sum_{\varphi\in\{h,v\}}$. Two like ($h-h$ and $v-v$) and one
     unlike ($v-h$) hard body interactions can be identified between
     $i$ and $j$ particles, which are given by
\begin{subequations}
\begin{align}
     u[(x,y,v),(x^\prime,y^\prime,v)]
 &=  \left\{ \begin{matrix}
              \infty & \text{if}\ |x-x^\prime|<D\\
              0      & \text{otherwise}
             \end{matrix}
     \right.
   ,
 \label{eq2}
 \\
     u[(x,y,h),(x^\prime,y^\prime,h)]
 &=  \left\{ \begin{matrix}
              \infty & \text{if}\ |x-x^\prime|<L\\
                     & \text{and} \ |y-y^\prime|<D\\
              0      & \text{otherwise}
             \end{matrix}
     \right. 
  ,
 \\
     u[(x,y,h),(x^\prime,y^\prime,h)]
 &=  \left\{ \begin{matrix}
              \infty & \text{if} \ |x-x^\prime|<\frac{L+D}{2}\\
              0      & \text{otherwise}
             \end{matrix}
     \right.
  .
 \label{eq4}
\end{align}
 \label{eq}
\end{subequations}
     Note that only the $h-h$ pair potential depends on the vertical
     positions, while the other two are the same for any $y$ and
     $y^\prime$ positions. This is due to the geometrical conditions, which
     allow only the horizontal particles to form two layers in the
     pore. After substitution of Eqs. (\ref{eq2})-(\ref{eq4}) into Eq.
     (\ref{eq1}) one can realize that the integrations in horizontal
     variables ($x_i$) cannot be achieved, i.e. the traditional
     transfer operator method cannot be applied for this model. However,
     on the basis of our previous study for parallel hard squares
     \cite{38}, it is feasible to work out a dimer-approach, where the
     two neighboring particles are considered as a dimer particle. If
     $N$ is an even number, i.e. we have $N/2$ dimers, the transfer
     operator method can be applied for the dimers, because as we show
     below the integrals in the dimer-dimer horizontal distances can
     be performed analytically. In the dimer-approach we introduce new
     notations. First of all, we label the dimers by capital letters
     to emphasize that the values of these indices are running up to
     $N/2$. Moreover, we introduce new variables instead of the
     original $x_i$ coordinates of the particles: $\hat{x}_I$ is the
     horizontal distance between the particles' centers of a dimer
     $I$, and $X_{I,J}$ is the horizontal distance between the centers
     of mass of dimers $I$ and $J$.  For the sake of simplicity we
     introduce
     $Q_I=(\hat{x}_I,y_{I,1},y_{I,2},\varphi_{I,1},\varphi_{I,2})$ for
     the inner coordinates of dimer $I$, where $y_{I,n}$ and
     $\varphi_{I,n}$ (here $n=1,2$) are only new notations for the $y$
     and $\varphi$ coordinates of the $2(I-1)+n$-th particle, which is
     the $n$-th member of the dimer $I$. At this point it is worth
     changing from the molecular coordinates to inner and relative
     neighboring ones ($Q_I$ and $X_{I,I+1}$) in Eq.  (\ref{eq1}) as
     follows $dL_x\prod_{i=1}^N dq_i\to dx_1 \prod_{I=1}^{N/2} dQ_I
     dX_{I,I+1}$ . With the new variables there is no need for $L_x$
     any more, as $L_x=\sum_{I=1}^{N/2}X_{I,I+1}$. However, the hard
     body exclusion sets the range of the inner variable $\hat{x}_I$
     in Eq.  (\ref{eq1}) as follows: $\hat{x}_I>\sigma_{I,1;I,2}$,
     where $\sigma_{I,1;I,2}$ is the horizontal contact distance
     between the first and second particles of the dimer
     $I$. One can write the horizontal contact distance
     generally between particles $i$ and $j$ as 
\begin{eqnarray}
      \sigma_{i;j}
   =  \left\{
       \begin{array}{ccl}
        0,            & \text{if} & \varphi_{i}=\varphi_{j}=h \text{ and } |y_{i}-y_{j}|>D\\
        L,            & \text{if} & \varphi_{i}=\varphi_{j}=h \text{ and } |y_{i}-y_{j}|<D\\

        D,            & \text{if} & \varphi_{i}=\varphi_{j}=v\\
        \frac{L+D}{2},& \text{if} & \varphi_{i} \neq \varphi_{j}
       \end{array}
      \right.
 \label{eq5}
\end{eqnarray}
     therefore the horizontal contact distance between particles $n$
     and $m$ of dimers $I$ and $J$ is
\begin{equation}
     \sigma_{I,n;J,m}
  =  \sigma_{2(I-1)+n,2(J-1)+m}
 .
\end{equation}
     Note that Eq. \ref{eq5} can be obtained easily from the pair
     interactions, see Eqs.~(\ref{eq}).  The lower bound of the
     neighboring dimer distance ($X_{i,i+1}$) is constrained by
     prohibiting the overlap between dimers. Therefore one can get
     that $\sigma_{I,I+1}<X_{I,I+1}$ in the partition function, where
     $\sigma_{I,I+1}$ is the horizontal contact distance between two
     neighboring dimers.  As first-first, second- first and
     second-second contacts may occur between particles of two
     neighboring dimers, the contact distance can be written as the
     maximum of the three possible values, as follows:
\begin{align}
      \sigma_{I,I+1}
   \equiv
       \max\left\{
           \sigma_{I,2;I+1,1} + \frac{\hat{x}_I + \hat{x}_{I+1}}{2},
          \right.
  \nonumber
 \\
          \left.
           \sigma_{I,1;I+1,1} + \frac{-\hat{x}_I + \hat{x}_{I+1}}{2},
          \right.
  \nonumber
 \\
          \left.
            \sigma_{I,2;I+1,2} + \frac{\hat{x}_I - \hat{x}_{I+1}}{2}
          \right\}
  .
 \label{eq:sigma_I,I+1}
\end{align}
     As can be seen from Eq.~\eqref{eq5}, $\sigma_{I,I+1}$ depends on
     the variables $Q_I$ and $Q_{I+1}$.  Using the new inner and outer
     variables one can rewrite Eq.  \ref{eq1} as follows 
\begin{eqnarray*}
     Z
 &=& \int \left( \prod_{I=1}^{N/2} dQ_I  dX_{I,I+1} \right)
      e^{-P \sum_{J=1}^{N/2} X_{J,J+1}}
  \nonumber
 \\
 &=& \int \left( \prod_{I=1}^{N/2} dQ_I \right)
     \prod_{J=1}^{N/2} \left( \int_{\sigma_{J,J+1}}^\infty dX_{J,J+1} 
      e^{-P X_{J,J+1}} \right)
,
\end{eqnarray*}
     where $P=\beta f_x$. The benefit of this form of $Z$ is that $N/2$
     integrations of $X_{I,I+1}$ can be performed analytically,
\begin{eqnarray}
     K_{I,I+1}=\int_{\sigma_{I,I+1}}^{\infty} dX_{I,I+1}e^{-PX_{I,I+1}}=\frac{e^{-P\sigma_{I,I+1}}}{P}.
\label{eq6}
\end{eqnarray}
     Using the definition of the continuum generalization of the matrix
     product, $K^2_{I,I+2}=\int dQ_{I+1} K_{I,I+1}K_{I+1,I+2}$, one
     can write the partition function very concisely in the following
     form 
\begin{eqnarray}
     Z
 &=& \int \left(\prod_{I=1}^{N/2} dQ_I \right) \prod_{J=1}^{N/2} K_{J,J+1}
  =  \int dQ_1 K_{1,1}^{N/2}
  \nonumber
 \\
 &=& \text{Tr} \left(\hat K^{N/2}\right)
 ,
\end{eqnarray}
     where Tr means the trace of the integral operator $\hat K$ defined
     by the kernel of Eq.~\eqref{eq6}. As the trace of an operator is
     independent of the used basis vectors, one can get $Z$
     straightforwardly in the eigenfunction frame, where
     $Z=\sum_l\lambda_l^{N/2}$ with $\lambda_l$ eigenvalues satisfying
     the following eigenvalue equation:
\begin{eqnarray}
     \int dQ_2 K_{1,2} \psi_l(Q_2)=\lambda_l\psi_l(Q_1),
\label{eq8}
\end{eqnarray}
     where $\psi_l(Q)$ is an eigenfunction of $\hat K$ corresponding to
     the eigenvalue $\lambda_l$. The variables of this function are
     the inner coordinates of a dimer. In the thermodynamic limit
     ($N\to\infty$) only the largest eigenvalue,
     $\lambda=\text{max}(\lambda_1,\lambda_2,\dots)$, contributes to
     the partition function, because $Z=\lim_{N\to\infty}\lambda^{N/2}
     \sum_l\left( \lambda_l/\lambda\right)^{N/2}=\lambda^{N/2}$. As
     $G=-k_BT \ln Z$, where $G$ is the Gibbs free energy, one can get
     that $\beta G/N=-(\ln \lambda)/2$. After straightforward but
     lengthy calculations one can show that the eigenfunction of Eq.
     (\ref{eq8}), using the notation
     $Q=(\hat{x},y,y',\varphi,\varphi^\prime)$ for the inner
     coordinates of a dimer, can be written as
\begin{widetext}
\begin{eqnarray}
     \psi(Q)
  =  \left\{
      \begin{array}{ll}
          \bar{\psi}(0,0) e^{-P\frac{\hat{x}-L+D}{2}} \;
          \theta\left(\frac{H-L}{2}-|y|\right) \;
          \theta\left(\frac{H-L}{2}-|y^\prime|\right) \;
          \theta(\hat{x}-D)
         &\text{for} \ \varphi=\varphi^\prime=v 
         \\
          \bar{\psi}(0,0) e^{-P\frac{\hat{x}-L+D}{2}}\;
          \theta\left(\frac{H-D}{2}-|y|\right)\;
          \theta\left(\frac{H-L}{2}-|y^\prime|\right)\;
          \theta\left(\hat{x}-\frac{L+D}{2}\right)
         &\text{for} \ \varphi=h,\varphi^\prime=v 
         \\
          \bar{\psi}(y^\prime,L) e^{-P\frac{\hat{x}-L}{2}}\;
          \theta\left(\frac{H-L}{2}-|y|\right)\;
          \theta\left(\frac{H-D}{2}-|y^\prime|\right)\;
          \theta\left(\hat{x}-\frac{L+D}{2}\right)
         & \text{for} \ \varphi=v,\varphi^\prime=h
         \\
          \left[\bar{\psi}(y^\prime,L) e^{-P\frac{\hat{x}-L}{2}}\;
            \theta(\hat{x}-L)+\bar{\psi}(y^\prime,\hat{x})
            \theta(L-\hat{x}) \theta(|y-y^\prime|-D)\right]\times
         &
         \\
         \mbox{}\hspace{1cm}
          \theta\left(\frac{H-D}{2}-|y|\right)\;
          \theta\left(\frac{H-D}{2}-|y^\prime|\right) 
         &\text{for} \ \varphi=\varphi^\prime=h 
      \end{array}
     \right.
\end{eqnarray}
     where $\psi(Q)$ denotes the eigenfunction of $\lambda$, $\theta(x)$
     is the Heaviside step function and $\bar{\psi}(y,\hat{x})$
     satisfies the following eigenvalue equation
\begin{eqnarray}
     \lambda\bar{\psi}(\hat{x}_1,y)
 &=& \frac{e^{-PL}}{P^2}\left[e^{-PD}(H-L)\bar{\psi}(0,0)
    +e^{-PL/2} \int_{-(H-D)/2}^{(H-D)/2}dy^\prime\bar{\psi}(L,y^\prime)\right] \times
  \nonumber
 \\
 &&\mbox{}\hspace{3.6cm}
     \left[\left(H-D-a(y)+(H-L) e^{P(L-D)}\right)e^{-P\hat{x}_1/2}
           +a(y)e^{P\hat{x}_1/2}
     \right]
  \nonumber
 \\
 && +\frac{e^{-PL}}{P}\int_{-(H-D)/2}^{(H-D)/2}dy^\prime 
       \left[A(y,y^\prime) 
         \left(e^{-P\hat{x}_1/2} \int_0^{\hat{x}_1}d\hat{x}_2 e^{P\hat{x}_2/2}
            \bar{\psi}(\hat{x}_2,y^\prime) + e^{P\hat{x}_1/2} 
            \int_{\hat{x}_1}^Ld\hat{x}_2e^{-P\hat{x}_2/2}
            \bar{\psi}(\hat{x}_2,y^\prime)
         \right)
        \right.
  \nonumber
 \\
 &&\mbox{}\hspace{3.6cm}
    +\left.\left(a(y^\prime)-A(y,y^\prime)\right)e^{-P\hat{x}_1/2} \int_0^Ld\hat{x}_2e^{-P\hat{x}_2/2}\bar{\psi}(\hat{x}_2,y^\prime)\right],
\label{eq10}
\end{eqnarray}
\end{widetext}
     where $a(y)$ is the available vertical distance for a free
     particle with orientation $h$ to pass another particle
     also with orientation $h$ fixed at the position $y$, while
     $A(y,y')$ is the same quantity, but there are two $h$ oriented
     fixed particles at $y$ and $y'$, see Ref.~\cite{38}. The
     probability distribution of the inner coordinates of a dimer is
     related to the eigenfunction of $\lambda$ eigenvalue through
     $f(Q)=\psi(Q)\psi^*(Q)$, where
     $\psi^*(Q)\equiv\psi^*(\hat{x},y,y^\prime,\varphi,\varphi^\prime)
     = \psi(\hat{x},y^\prime,y,\varphi^\prime,\varphi)$. As $f(Q)$ is
     normalized, i.e. $\int dQ f(Q)=1$, the one particle distribution
     function is given by $f(y,\varphi) = \int d\hat{x}\,
     dy^\prime\sum_{\varphi^\prime}\,
     f(\hat{x},y,y^\prime,\varphi,\varphi^\prime)$, and the fraction
     of particles along the horizontal and vertical direction can be
     obtained from $x_\varphi=\int dy\, f(y,\varphi)$.

     In lack of the second neighbor interactions, which happens for
     $H\leq 2D$, the transfer operator method becomes much simpler as
     only the second particle of the first dimer and the first
     particle of the second dimer can get in contact, i.e.
     Eq.~\eqref{eq:sigma_I,I+1} simplifies to
\begin{eqnarray}
     \sigma_{I,I+1}
  =  \sigma_{I,2;I+1,1} + \frac{\hat{x}_I+\hat{x}_{I+1}}{2}.
\end{eqnarray}
     Moreover, $\sigma_{I,2;I+1,1}$ depends only on the orientations of
     particles but not on $y$.  Substituting this form into the
     eigenvalue equation Eq~\eqref{eq8} one can found that the
     eigenfunction must have a much simpler form than it was in the
     case of wider pore: $\psi(Q)=\psi_{\varphi^\prime} e^{-P\hat
       x/2}$. Therefore the integrals over all the positional
     variables can be performed analitycally, and finally one can see
     that the kernel of Eq.~\eqref{eq6} becomes a product and the
     eigenvalue equation reduces to a simple system of linear equation
     with a discrete $2 \times 2$ matrix kernel:
\begin{equation}
     \sum_{\varphi^\prime} 
         (H-\sigma(\varphi^\prime)) \tilde K_{\varphi,\varphi^\prime} \psi_{\varphi^\prime}
  =  \tilde\lambda \psi_\varphi
 ,
 \label{eq:eigen_narrow}
\end{equation}
     where $\tilde\lambda^2 = \lambda$,
     $\tilde{K}_{\varphi,\varphi^\prime} =
     P^{-1}e^{-P\sigma(\varphi,\varphi^\prime)}$ and
     $\sigma(\varphi,\varphi^\prime)$ is the horizontal contact
     distance between neighbouring particles which, as can be seen
     from Eq.~\eqref{eq5}, depends only on the orientation as follows:
     $\sigma(h,h)=L$, $\sigma(v,v)=D$,
     $\sigma(v,h)=\sigma(h,v)=(L+D)/2$. We remind here that
     $\sigma(\varphi)$ is the vertical length of the particle. In this
     case we get the Gibbs free energy from $\beta G/N=-\ln
     \tilde{\lambda}$.  

     Having obtained the Gibbs free energy from the solution of the
     eigenvalue equations \eqref{eq10},\eqref{eq:eigen_narrow}, one
     can get several properties from the standard thermodynamic
     relations. The average horizontal dimension of the pore ($\langle
     L_x\rangle$) and the vertical force are given by
     $\displaystyle{\langle L_x\rangle= \frac{\partial G}{\partial
         f_x}}$ and $\displaystyle{f_y=-\frac{\partial G}{\partial
         H}}$ as $dG=\langle L_x\rangle df_x-f_y dH$ at fixed $T$ and
     $N$. The number density of the system ($\rho=N/A$, where
     $A=\langle L_x\rangle H$) is simple coming from
     $\displaystyle{\rho^{-1}=H\frac{\partial G/N}{\partial f_x}}$.
     In practice, it is better to use the horizontal and vertical
     pressures ($P_x,P_y$), which are defined as $P_x=f_x/H$ and
     $P_y=f_y/L_x$. Using $P_x$ and $P_y$ one can write that
     $\displaystyle{\rho^{-1}=\frac{\partial G/N}{\partial P_x}}$ and
     $\displaystyle{P_y=-\rho H\frac{\partial G/N}{\partial H}}$. It
     is also possible to determine the isothermal compressibility
     ($\chi_T$) and the isobaric heat capacity ($C_P$) from
     $\displaystyle{\chi_T=-\frac{1}{A}\frac{\partial A}{\partial
         P_x}}$ and $\displaystyle{C_P=-P_x\frac{\partial A}{\partial
         T}}$. We use $D$ as the unit of length and present all
     quantities in dimensionless units. These are the followings:
     $H^*=H/D$, $L^*=L/D$, $\eta=\rho LD$, $P_x^*=\beta P_xLD$,
     $P_y^*=\beta P_yLD$,
     $\displaystyle{\chi_T^*=k_BT\frac{\chi_T}{LD}}$ and
     $\displaystyle{C_P^*=\frac{C_P}{Nk_B}}$.

     \section{Results}

     The wall-particle interactions favour planar ordering at the
     wall, while the particle- particle interaction promotes the long
     axes of the rods to be parallel with each other. The competition
     of these ordering effects and the packing constrain give rise to
     three different structures: (i) a low density one-layer fluid
     phase with planar ordering (FH1), (ii) a fluid with two layers
     and planar ordering (FH2) and (iii) a fluid with one layer and
     homeotropic ordering (FV). The appearance of these structures
     with increasing horizontal pressure is shown in Fig.  \ref{fig1}.

     We start by presenting our analytical results for $H^*\leq 2$,
     where only first neighbor interactions are present, i.e. the
     rectangles are not allowed to pass each other in all possible
     configurations.  It can be shown easily that the largest
     eigenvalue of Eq.  (\ref{eq:eigen_narrow}) is
     $\tilde{\lambda}=a+b$, where $a=(H-L)\tilde{K}_{v,v}$ and
     $b=(H-D)\tilde{K}_{h,h}$. From Eq.~\eqref{eq:eigen_narrow} and
     $\psi(Q)=\psi_{\varphi^\prime} e^{-P\hat x/2}$ we can get the
     one-particle distribution function which is given by
     $\displaystyle{f(y,h) =\frac{b}{(a+b)(H-D)}}$ and
     $\displaystyle{f(y,v)=\frac{a}{(a+b)(H-L)}}$. Note that $f$ is
     independent of $y$. Furthermore, the mole fractions for
     horizontal and vertical orientations can be obtained from
     $x_h=b/(a+b)$ and $x_v=a/(a+b)$ satisfying the normalization
     condition $x_h+x_v=1$. From $\beta G/N=-\ln \tilde{\lambda}$, we
     are in a position to get analytical results for the packing
     fraction ($\eta$), vertical pressure ($P_y$), isothermal
     compressibility ($\chi_T$) and isobaric heat capacity $(C_P)$.
     The results are summarized in the following equations
\begin{eqnarray}
&&\eta=\frac{LD}{H}\left(\frac{1}{P}+\frac{aD+bL}{a+b}\right)^{-1},\label{eq13}\\
&&P_y^*=\frac{\eta H}{a+b}\left(\frac{a}{H-L}+\frac{b}{H-D}\right),\label{eq14}\\
&&\chi_T^*=\eta \left(\frac{H}{LD}\right)^2\left(\frac{1}{P^2}+\frac{aD^2+bL^2}{a+b}
-\left(\frac{aD+bL}{a+b}\right)^2\right),\nonumber\\
&&\label{eq16}\\
&&C^*_P=\frac{\left(P_x^*\right)^2}{\eta}\chi_T^*.\label{eq17}
\end{eqnarray}
     Note that the same results can be obtained from Eq. (\ref{eq10})
     and $\beta G/N=-(\ln\lambda)/2$, because $a(y_1)=0$ and
     $A(y_1,y_2)=0$ for $H^*\leq 2$.
\begin{figure}
\includegraphics[width=8cm]{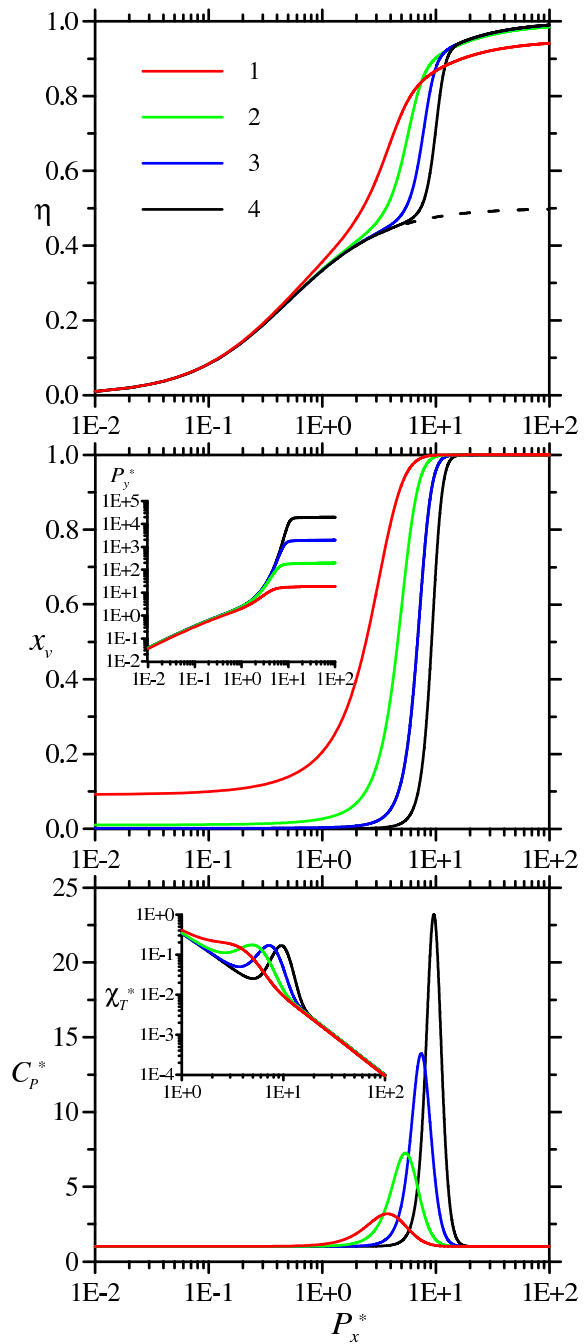}
\caption{ Orientational ordering transition of hard rectangles for
  pore-width $H^*=2$. The length of the rectangle is varied via
  $L^*=2-10^{-\epsilon}$, where $\epsilon=1$, 2, 3 and 4.  The packing
  fraction ($\eta$), the mole fraction of vertically oriented
  particles ($x_v$) and the isobaric heat capacity ($C_p$) are shown
  as a function of horizontal pressure ($P_x$) in the upper, middle
  and lower panels, respectively. The dashed curve in the upper panel
  is the Tonks-equation of horizontally ordered hard rectangles. The
  vertical pressure ($P_y$) and the isothermal compressibility
  ($\chi_T$) are also shown in the insets of middle and lower panels.
  All quantities are in dimensionless unit, i.e. $H^*=H/D$, $L^*=L/D$,
  $P_x^*=\beta P_xLD$, $P_y^*=\beta P_yLD$, $\chi_T^*=k_BT\chi_T/(LD)$
  and $C_P^*=C_p/Nk_B$.  }
\label{fig2}
\end{figure}

     Fig. \ref{fig2} presents together the results of Eqs.
     (\ref{eq13})-(\ref{eq17}) for varying $L$ at $H^*=2$. One can see
     that a structural transition occurs between horizontally and
     vertically ordered fluids.  The transition becomes more
     pronounced as $L^*=2-10^{-\epsilon}$ goes to $H^*$ with
     increasing $\epsilon$. To understand this phenomenon, we examine
     the low, intermediate and high-pressure cases as obtained from
     the equation of state [Eq. (\ref{eq13})]. At very low pressures
     the interaction term $(aD+bL)/(a+b)$ in Eq. (\ref{eq13}) becomes
     negligible with respect to $P^{-1}$ and the ideal gas law
     ($P_x^*=\eta$) can be reproduced, i.e. all curves go together in
     the $\eta-P_x^*$ diagram for $P_x^*<1$. At intermediate
     pressures, $b$ becomes much larger than $a$ with increasing
     $\epsilon$, i.e.  $(aD+BL)/(a+b)\approx L$. From this fact we get
     that
\begin{eqnarray}
\eta=LD\left[H\left(\frac{1}{P}+L\right)\right]^{-1},
\label{eq18}
\end{eqnarray}
     which is the well-known Tonks-equation of horizontally ordered
     rectangles. The close packing density of this fluid can be
     obtained in the $P\to\infty$ limit, and we get that
     $\eta_{cp}=D/H$ ($\eta_{cp}=1/2$ in Fig. \ref{fig2}). One can see
     that the perfect agreement between Eqs. (\ref{eq13}) and
     (\ref{eq18}) extends to the direction of higher pressures with
     increasing $\epsilon$.  The reason for this is that the vertical
     accessible distance and the corresponding fluctuations in $y$
     positions shrink with $L\to H$ for particles in vertical
     direction, while particles can occupy the same vertical interval
     and $y$ fluctuations are not affected as $L\to H$ for particles
     with horizontal direction. Practically this manifests in
     prefactors of $a$ and $b$, where ($H-L$) and ($H-D$) are the
     accessible distances along the $y$ direction. Therefore $b$ is
     the dominant factor for $L\to H$.  However, this is not the case
     for very high pressures, because we reach the close packing at
     $\eta=0.5$ with the horizontally ordered particles, while the
     vertically ordered fluid has the maximum density at $\eta=L/H$ ,
     which is almost one in our case. Therefore the system of
     rectangles must undergo a structural transition for $\eta>0.5$.
     In this case $a$ overcomes $b$ and $(aD+BL)/(a+b)\approx D$,
     because $\tilde{K}_{v,v}$ becomes much larger than
     $\tilde{K}_{h,h}$ with $P\to\infty$. In this case the system
     becomes a fluid of vertically ordered rectangles, which can be
     described by the following Tonks-equation
\begin{eqnarray}
\eta=LD\left[H\left(\frac{1}{P}+D\right)\right]^{-1},
\end{eqnarray}
     which gives the close packing density ($\eta=L/H$) in the limit
     $P\to\infty$. Practically the suppression of horizontal
     fluctuations is the driving force of the structural transition,
     which manifests in a crossing between $a$ and $b$ interaction
     terms. The structural change can be seen very clearly in the mole
     fraction, too, because $x_v\approx 0$ for intermediate pressures
     ($b>a$), while $x_v\approx 1$ for high pressures ($b<a$). The
     heat capacity is the same in the low and high pressure limit as
     $C_P^*(P\to 0)=C_P^*(P\to\infty)=1$, but it exhibits a peak at
     the transition region.  As the transition becomes sharper ($L\to
     H$), the peak is narrower and higher. The structural transition
     has a signature in the compressibility factor, too, since the
     system can be compressed more in the transition region than in
     the outer region.  It can be seen from Eq. (\ref{eq16}) that the
     system becomes incompressible with increasing pressure as
     $\chi_T^*(P\to 0)=0$. The vertical pressure increases with $P$ in
     the horizontal fluid phase, while it saturates in the vertical
     fluid one. Its maximum value is given by
     $P_y^*(P\to\infty)=L/(H-L)$, which can be obtained from Eq.
     (\ref{eq14}) using $a\gg b$ and $\eta\to L/H$.

\begin{figure}
\includegraphics[width=8cm]{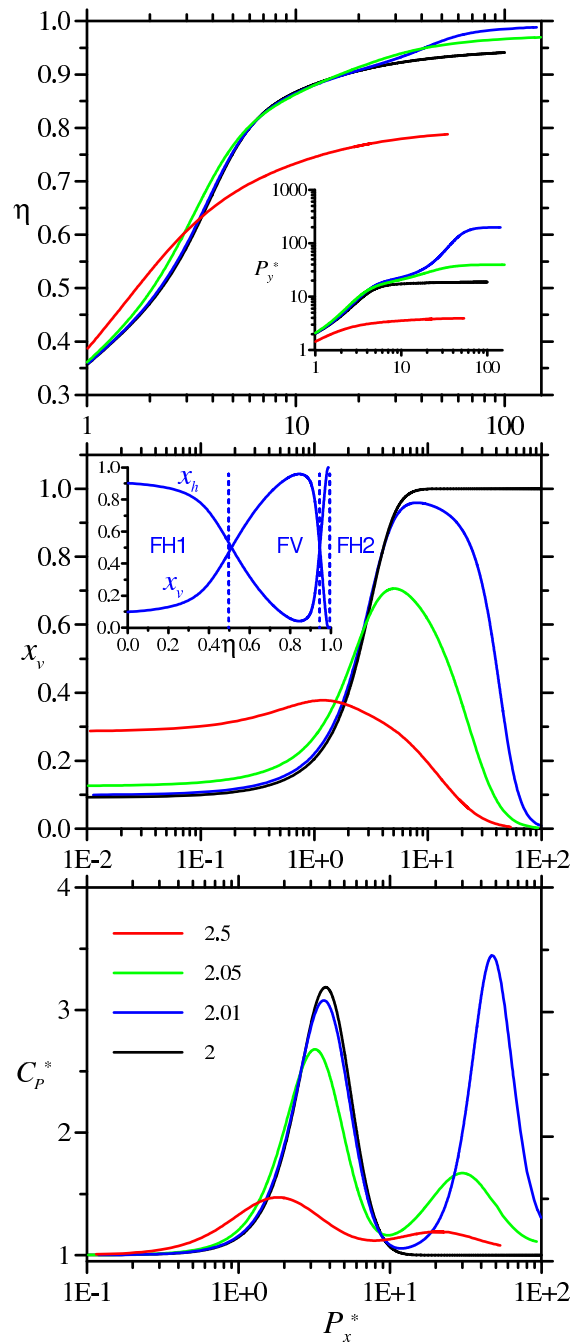}
\caption{ Positional and orientational ordering transitions of hard
  rectangles at $L^*=1.9$. The width of the pore is varied from
  $H^*=2$ to $H^*=2.5$. The packing fraction ($\eta$), the mole
  fraction of vertically oriented particles ($x_v$) and the isobaric
  heat capacity ($C_p$) as a function of horizontal pressure are shown
  in the upper, middle and lower panels, respectively. The mole
  fractions of vertically ($x_v$) and horizontally ($x_h$) ordered
  rectangles as a function of packing fraction is shown in the inset
  of the middle panel, while the vertical pressure ($P_y$) is
  presented in the inset of upper panel. The vertical dashed lines are
  the packing fractions of the close packing structures in FH1, FV and
  FH2 phases, respectively. All quantities are in dimensionless unit.
}
\label{fig3}
\end{figure}

\begin{figure}
\includegraphics[width=8cm]{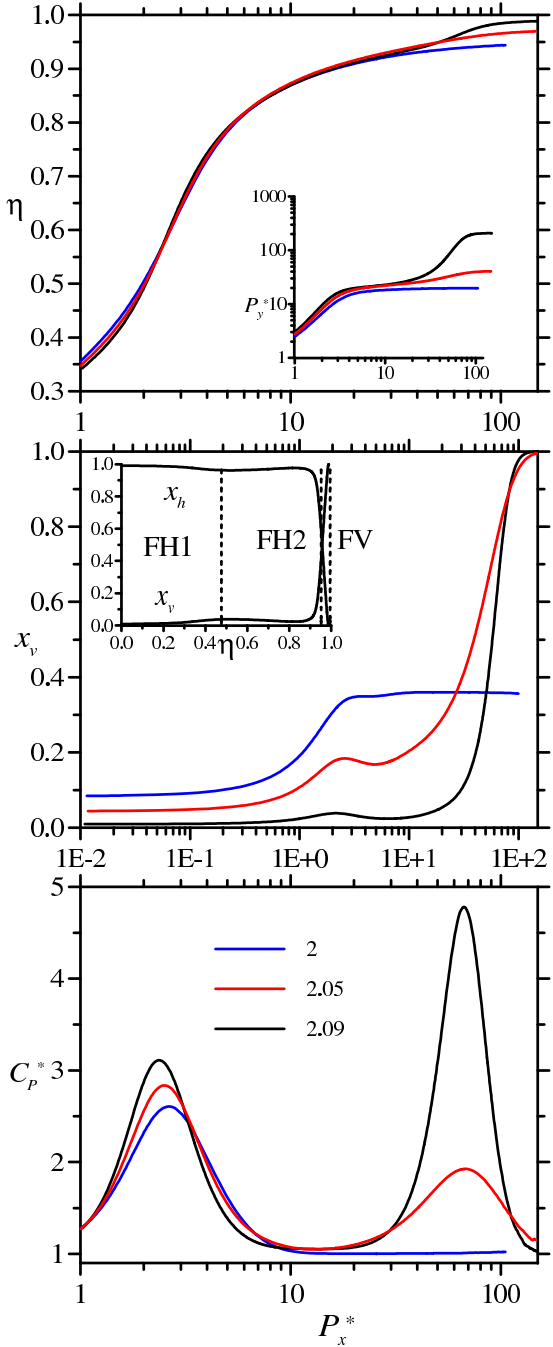}
\caption{ Same as Fig. \ref{fig3} for pore-width $H^*=2.1$. The length
  of the rectangle is varied from $L^*=2$ to $L^*=2.09$.  }
\label{fig4}
\end{figure}

     In wider pores ($H^*>2$), where the horizontal particles can pass
     each other, the close packing structure and the corresponding
     volume fractions depend on the length of the particle. For $L<2D$
     the two layer structure (FH2) is the most dense with
     $\eta_{cp}=2D/H$, while FV has the highest close packing for
     $L>2D$ with $\eta_{cp}=L/H$. Fig.  \ref{fig3} shows how the FH2
     structure conquers the high density region as the pore is
     widening from $H^*=2$ to $H^*=2.5$ at $L^*=1.9$.  One can observe
     two structural transitions with increasing horizontal pressure: a
     FH1-FV transition occurs when the packing fraction exceeds the
     close packing of the FH1 structure ($\eta_{cp}(\rm FH1)=D/H$),
     while the FV-FH2 transition takes place at the vicinity of
     $\eta\approx \eta_{cp}(\rm FV)=L/H$. This manifests in two
     inflection points in the $\eta-P_x^*$ curve, two plateaus in
     $P_y^*$ and two peaks in $C_P^*$. The stabilization of the FH1
     phase at low densities is due to the wall, because the pore is
     wider for particles having horizontal orientation than for those
     with vertical orientation, i.e.  a larger portion of the pore can
     be occupied with the FH1 structure.  As the close packing of FH1
     is approached, the system must change its structure to avoid the
     forbidden overlapping states. It chooses the FV structure,
     because it provides plenty of room along the $x$ axis with a
     price of less room along $y$ axis. This is entropically better
     than going directly to the FH2 structure, where the room for the
     particles is less both in $x$ and $y$ directions.  However this
     is not the case in the vicinity of the close packing density of
     the FV phase, because particles get in contact and are forced to
     choose the more packed FH2 structure, where the particles still
     have some room along the $x$ and $y$ axes. The FH1-FV and FV-FH2
     transitions are getting sharper as $L\to 2D$ ($L<2D$) and $H\to
     2D$ ($H>2D$), while they become smoother with increasing pore
     width.  For example the appearance of mixed phases at low and
     intermediate densities for $H^*=2.5$ is due to the increased room
     available along the horizontal direction for both vertical and
     two-layer structures. The FH1-FV-FH2 phase sequence is replaced
     by FH1-FH2-FV if $L>2D$, which is displayed in Fig \ref{fig4}.
     At $L=2D$ only a structural transition occurs between FH1 and a
     mixed phase with more particles in FH2 structure.  The emergence
     of this mixed phase is due to the fact that the close packing
     densities of FH1 and FV are identical at $L=2D$ ($\eta_{cp}(\rm
     FV)=\eta_{cp}(\rm FH2)$). This manifest clearly in the one-peak
     structure of $C_P^*$, which is located in the vicinity of the
     close packing density of the FH1 phase. With gradual increment of
     $L$ to $H$ one can see the emergence of a new inflection point in
     the $\eta-P_x^*$ curve, a second plateau in $P_y^*$ and a second
     peak in $C_P^*$ in the vicinity of $\eta\approx \eta_{cp}(\rm
     FH2)=2D/H$.  The transition is getting sharper and the phases
     become less mixed as $L\to H$, because the vertical fluctuations
     are suppressed.

\begin{figure}
\includegraphics[width=8cm]{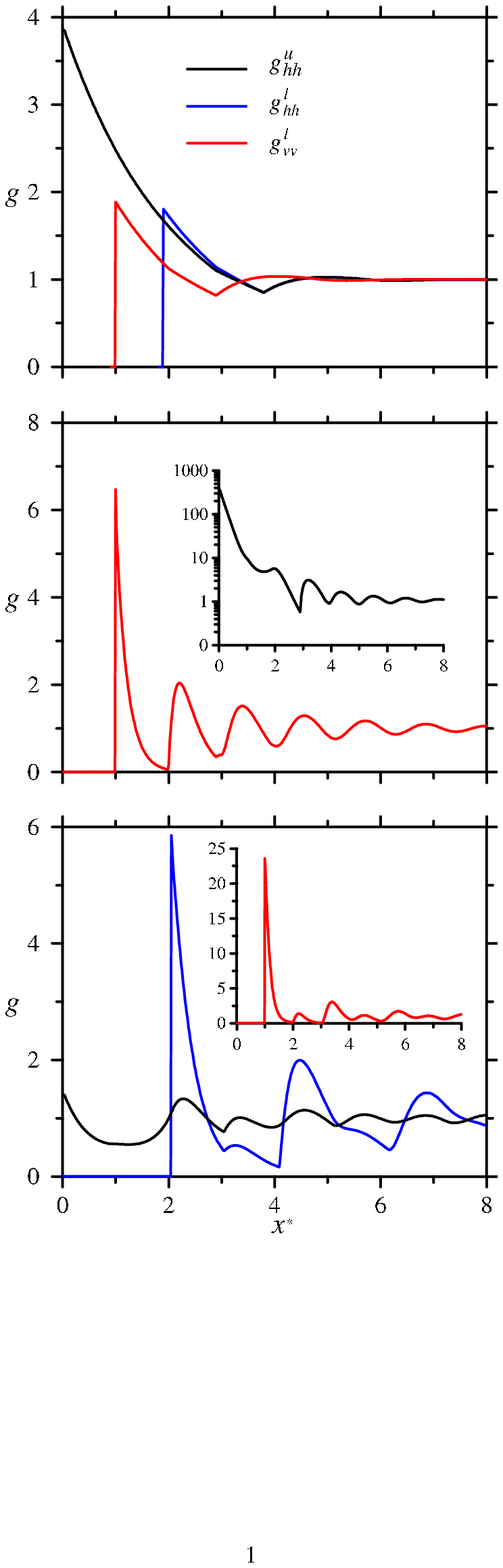}
\caption{ Pair correlation functions of FH1 (upper panel), FV (middle
  panel) and FH2 (lower panel) structures. Three different functions
  are shown for a particle fixed at the lower wall: in $g_{hh}^l(x)$
  [$g_{hh}^u(x)$] both particles are horizontal and the second one is
  at the lower [upper] wall, while in $g_{vv}^l(x)$ both particles are
  vertical and the second one is at the lower wall. These results are
  obtained with the FMT and the following parameters: $H^*=2.05$ ,
  $L^*=1.9$ and $\eta=0.2$ (upper panel), $H^*=2.05$ , $L^*=1.9$ and
  $\eta=0.68$ (middle panel) and $H^*=2.1$, $L^*=2.05$ and $\eta=0.8$
  (lower panel).  }
\label{fig5}
\end{figure}

     Finally we present our fundamental measure theory (FMT) results
     for the structural properties of the three different structures
     observed.  Here we do not go into the details of the theory, as
     it can be found in \cite{39}. All correlation functions are
     computed for a fixed particle in contact with the lower wall. The
     following pair correlation functions are determined as a function
     of horizontal distance $(x)$: in $g_{hh}^l(x)$ both particles are
     in horizontal orientation and are located at the contact with the
     lower wall, in $g_{hh}^u(x)$ both particles are in horizontal
     orientation and the second particle is at the upper wall, while
     in $g_{vv}^l(x)$ and $g_{vv}^u(x)$ both particles are in vertical
     orientation.  The typical correlation functions of the FH1, FH2
     and FV structures are shown together in Fig. \ref{fig5}.  At
     $H^*=2.05$, $L^*=1.9$ and $\eta=0.2$ we find a FH1 structure, as
     FMT gives $x_h=0.84$ ($x_v=0.16$). One can see that this is a
     weakly correlated system as all correlation functions go quickly
     to 1. The correlation between two horizontal particles is maximum
     when they are one above the other (see $g_{hh}^u(x)$). However
     this configuration happens rarely because it is more common that
     the horizontal particles are in the same layer.  Note that
     $g_{vv}^l(x)$ and $g_{vv}^u(x)$ are the same in this case, and
     show very small correlations at short distances. At $\eta=0.68$
     the stable structure is FV, because $x_v=0.84$ and $x_h=0.16$.
     The correlation between vertical particles at both walls is
     identical.  $g_{vv}^l(x)$ shows the typical fluid structure of
     vertically ordered particles as the period of the damped
     oscillation is close to $D$. The correlations between vertical
     particles are relatively high when the particles are in contact
     and decay with distance. The behavior of $g_{hh}^u(x)$ gives
     further information about the microstructure of the FV structure
     for wider pores ($H>2D$).  One can see an extremely strong
     correlation for very short distances between horizontal particles
     (see the log scale in the graph). This is due to the fact that
     the density is very high and almost all particles are vertical.
     To maintain the high density of the system the horizontal
     particles must form dimers. For larger distances the correlations
     decay to 1 with some oscillations.  The correlations in the same
     layer ($g_{hh}^l(x)$) have a similar behavior but with a
     forbidden region and with the first peak much smaller than
     $g_{hh}^u(0)$. These results show very clearly that the FV
     structure of wide pores consists of long chains of vertical
     particles which are interrupted mainly by the dimers of
     horizontal particles. This is not the case in narrow pores
     ($H<2D$), where dimers cannot form along the $y$ direction.  The
     structure of FH2 is examined at $H^*=2.1$, $L^*=2.05$ and
     $\eta=0.8$, where $x_h=0.93$ ($x_v=0.07$).  Here $g_{hh}^u(x)$
     shows very clearly that the two layers of horizontal particles
     are quite uncorrelated since $g_{hh}^u(x)$ oscillates always very
     close to 1.  However in the same layer $g_{hh}^l(x)$ shows that
     the horizontal particles are more correlated with a period close
     to $L$. In both $g_{hh}^l(x)$ and $g_{hh}^u(x)$ there exist two
     kinds of peaks, the main ones have a period of $\sim 2L$, while
     the secondary smaller peaks are dephased by $\sim D$ due to the
     presence of few vertical particles.  The correlation between
     vertical particles ($g_{vv}(x)$) shows a first peak at $x=D$
     (particles at contact) which is significantly higher than the
     rest of the peaks. This means that the vertical particles can be
     found mainly in the form of dimers in the sea of horizontal
     particles. As all correlations are short ranged, FH2 is really a
     fluid phase without signs of solid-like order.

     \section{Discussion}

     We have shown that the structure of the rectangles can be
     manipulated in slit-like pore by an external force acting along
     the longitudinal direction. Upon increment of the force the
     particles undergo one or two structural rearrangements. One or
     two layers can form with horizontally ordered particles, while
     only one layer can be generated with vertically ordered ones. As
     the thermodynamic quantities do not exhibit singularities, the
     possibility of a genuine phase transition can be excluded in our
     model. This is a consequence of the form of the integral kernel
     (Eq. \ref{eq6}), which is positive for arbitrary pressure and
     molecular parameters \cite{40,41}. However, the inflection points
     of the equation of state, plateaus in vertical pressure and peaks
     in the heat capacities are hallmarks of the structural phase
     transitions. As the system can be trapped easily by the
     suppression of room available for the particles along vertical
     direction, the system can get into jammed and glass states as
     happens with hard disks confined into a 2D slit-pore
     \cite{42,43,44}. In addition to this, the very strong structural
     transition as occurs in our model in the limit of $L\to H$ can be
     viewed as a first order transition in simulation studies
     \cite{45}. The advantage of our TOM formalism is that it provides
     exact thermodynamic results and some structural information.

     It is particularly interesting in this model that anchoring of
     the particles from planar to hometropic takes place with
     increasing longitudinal pressure or density and the planar
     ordering effect of the hard walls does not prevail at high
     pressures. Recently a density induced planar to homeotropic
     ordering was observed at hard walls in the system of stiff ring
     polymers \cite{46}. Therefore it is feasible that strongly
     confined hard rods can also exhibit planar- homeotropic ordering
     transition if the length of the rods is close to the pore width.
     In this regard further studies are needed.

     \acknowledgments

     Financial support from MINECO (Spain) under grants
     FIS2013-47350-C5-1-R and FIS2015-66523-P are acknowledged.


\begin{references}
\bibitem{1} M. Schmidt and H. L\"owen, PRE 55, 7228 (1997).
\bibitem{2} A. Fortini and M. Dijkstra, J. Phys.: Condens. Matter 18, L371 (2006).
\bibitem{3}R. van Roij, M. Dijkstra, and R. Evans, EPL 49, 350 (2000).
\bibitem{4} M. Dijkstra, R. van Roij, and R. Evans, PRE 63, 051703 (2001).
\bibitem{5} R. van Roij, M. Dijkstra, and R. Evans, J. Chem. Phys. 113, 7689 (2000).
\bibitem{6} M. C. Lagomarsino, M. Dogterom, and M. Dijkstra, J. Chem. Phys. 119, 3535 (2003).
\bibitem{7} R. Aliabadi, M. Moradi and S. Varga, PRE 92, 032503 (2015).
\bibitem{8} D. de las Heras, E. Velasco, and L. Mederos, PRL 94, 017801 (2005).
\bibitem{9} D. de las Heras, E. Velasco, and L. Mederos, PRE 74, 011709 (2006).
\bibitem{10} K. W. Wojciechowski and D. Frenkel, Comp. Methods in Science and Tech. 10, 235
(2004).
\bibitem{11} A. Donev, J. Burton, F. H. Stillinger and S. Torquato, PRB 73, 054109 (2006).
\bibitem{12} C. Avendano and F. A. Escobedo, Soft Matter 8, 4675 (2012).
\bibitem{13} H. Schlacken, H.-J. Mogel, and P. Schiller, Mol. Phys. 93, 777 (1998).
\bibitem{14} S. Varga and I. Szalai, J. Mol. Liqs. 85, 11 (2000).
\bibitem{15} Y. Mart\'{\i}nez-Rat\'on, E. Velasco, and L. Mederos, J. Chem. Phys. 122, 064903 (2005).
\bibitem{16} Y. Mart\'{\i}nez-Rat\'on, E. Velasco, and L. Mederos, J. Chem. Phys. 125, 014501 (2006).
\bibitem{17} S. Belli, M. Dijkstra and R. van Roij, J. Chem. Phys. 137, 124506 (2012).
\bibitem{18} J. Kundu and R. Rajesh, PRE 89, 052124 (2014).
\bibitem{19} T. Nath, D. Dhar and R. Rajesh, EPL 114, 10003 (2016).
\bibitem{20} J. Galanis, D. Harries, D. L. Sackett, W. Losert and R. Nossal, PRL 96, 028002 (2006).
\bibitem{21} R. S. Mclean, X. Huang, C. Khripin, A. Jagota and M. Zheng, Nano Letters 6, 55 (2006).
\bibitem{22} K. Zhao, C. Harrison, D. Huse, W. B. Russel and P. M. Chaikin, PRE 76, 040401 (2007).
\bibitem{23} K. Zhao, R. Bruinsma, and T. G. Mason, Proc. Natl. Acad. Sci. U.S.A. 108, 2684 (2011).
\bibitem{24} R. S\'anchez and L. A. Aguirre-Manzo, Phys. Scr. 90, 095002 (2015).
\bibitem{25} L. Walsh and N. Menon, J. Stat. Mech.: Theory and Experiment 2016, 083302 (2016).
\bibitem{26} W.-Y. Zhang, Y. Jiang, and J. Z. Y. Chen, PRL 108, 057801 (2012).
\bibitem{27} J. Z. Y. Chen, Soft Matter 9, 10921 (2013).
\bibitem{28} M. Gonz\'alez-Pinto, Y. Mart\'{\i}nez-Rat\'on, and E. Velasco, PRE 88, 032506 (2013).
\bibitem{29} M. E. Ferraro, T. M. Truskett and R. T. Bonnecaze, PRE 93, 032606 (2016).
\bibitem{30} Y. Li, H. Miao, H. Ma, and J. Z. Y. Chen, Soft Matter 9, 11461 (2013).
\bibitem{31} D. de las Heras and E. Velasco, Soft Matter 10, 1758 (2014).
\bibitem{32} T. Geigenfeind, S. Rosenzweig, M. Schmidt, and D. de las Heras, J. Chem. Phys. 142,
174701 (2015).
\bibitem{33} Y. Mart\'{\i}nez-Rat\'on, PRE 75, 051708 (2007).
\bibitem{34} D. A. Triplett and K. A. Fichthorn, PRE 77, 011707 (2008).
\bibitem{35} J. L. Lebowitz, J. K. Percus, and J. Talbot, J. Stat. Phys. 49, 1221 (1987).
\bibitem{36} D. A. Kofke and A. J. Post, J. Chem. Phys. 98, 4853 (1993).
\bibitem{37} Y. Kantor and M. Kardar, EPL 87, 60002 (2009).
\bibitem{38} P. Gurin and S. Varga, J. Chem. Phys. 142, 224503 (2015).
\bibitem{39} M. Gonz\'alez-Pinto, Y. Mart\'{\i}nez-Rat\'on, S. Varga, P. Gurin and E. Velasco, 
J. Phys.: Condens. Matter 28, 244002 (2016).
\bibitem{40} L. van Hove, Physica 16, 137 (1950).
\bibitem{41} J. A. Cuesta and A. S\'anchez, J. Stat. Phys. 115, 869 (2004).
\bibitem{42} S. S. Ashwin and R. K. Bowles, PRL 102, 235701 (2009).
\bibitem{43} M. J. Godfrey and M. A. Moore, PRE 89, 032111 (2014).
\bibitem{44} M. J. Godfrey and M. A. Moore, PRE 91,022120 (2015).
\bibitem{45} P. Gurin, S. Varga and G. Odriozola, PRE 94, 050603(R) (2016).
\bibitem{46} P. Poier, S. A. Egorov, C. N. Likos and R. Blaak, Soft Matter 12, 7983 (2016).
\end{references}
\end{document}